\documentclass[aps,twocolumn,showpacs,superscriptaddress,groupedaddress]{revtex4}  
\usepackage{graphicx}  
\usepackage{dcolumn}   
\usepackage{bm}        
\usepackage{amssymb}   

\usepackage{wasysym} 

\begin{document}



\title{The Paleoclimatic evidence for  Strongly Interacting Dark Matter \\
Present in the Galactic Disk}

\author{Nir J. Shaviv}
\address{Racah Institute of Physics, Hebrew University of Jerusalem, Jerusalem 91904, Israel}

\def\ddm{\mathrm{dDM}}
\def\hdm{\mathrm{hDM}}
\def\kpc{\mathrm{kpc}}

\begin{abstract}
 Using a recent geochemical reconstruction of the Phanerozoic climate which exhibits a 32 Ma oscillation with  a phase and the secondary modulation expected from the vertical the motion of the solar system perpendicular to the galactic plane \cite{shaviv:2014}, we show that a kinematically cold strongly interacting disk dark matter (dDM) component is necessarily present in the disk. It has a local density $\rho_\mathrm{\ddm} = 0.11 \pm 0.03$~M$_{\odot}/$pc$^3$. It is also consistent with the observed constraints on the total gravitating mass and  the baryonic components, and it is  the natural value borne from the Toomre stability criterion. It also has surface density $\Sigma_{\ddm} = 15 \pm 5$~M$_{\odot}/$pc$^2$ and a vertical velocity dispersion of $\sigma_{W} = 8.0 \pm 4.5$~km/s. A dense (``dinosaur killing") thin disk is ruled out. The ``normal" halo dark matter (hDM) component should then have a local density $\rho_\mathrm{hDM} \lesssim 0.01$~M$_{\odot}/$pc$^3$. If the dDM component follows the baryons, its average density parameter is $\Omega_{\ddm} = 1.5 \pm 0.5\%$ and it comprises about 1/8 to 1/4 of Milky Way (MW) mass within the solar circle. 
\end{abstract}

\pacs{
98.35.Ce,	
98.35.Df,	
98.35.Hj,	
95.35.+d  
}
\maketitle

\def\secskip{\vskip -8pt}
\section{Introduction}
\secskip

A standard astronomical method to indirectly detect dark matter in the MW disk is to find a difference between kinematic determinations of the total density of gravitating mass and estimates for the baryonic mass density. This type of evidence for missing mass can be traced to Oort \cite{oort:1932,oort:1960}. The local mass density measurement itself,  known as the ``Oort limit", was found by him to be $\rho \sim 0.15$~M$_{\odot}/$pc$^3$, more than the observed baryon density. Subsequent analyses were contradictory. Whereas some recovered Oort's result \cite{bahcall:1984,bahcall:1992}, others found no conclusive evidence for any missing matter \cite{bienayme:1987,kuijken:1989,kuijken:1991}. 
 The debate was mostly considered resolved with the analysis of the Hipparcos data \cite{holmberg:2000}, giving $\rho_\mathrm{total} \sim 0.102\pm 0.01 $~M$_{\odot}/$pc$^3$, compared with $\rho_\mathrm{baryon} \sim 0.095$~M$_{\odot}/$pc$^3$. 
 
 The  difference is consistent with the small amount of dark matter expected from the ``halo" dark matter (hDM) component. Extrapolating the dark matter density from $z=1- 4$ kpc  to the plane gives $\rho_{\hdm} = 0.008 \pm 0.003~$M$_\odot/$pc${^3}$ \cite{bovy:2012}. Similarly, a standard spherically symmetric NFW profile that would fit the rotation curve at the solar galactic radius gives $\rho_{\hdm} = 0.0084~$M$_\odot/$pc${^3}$ \cite{Olling:2001}.
Thus, measurements of the different densities at the plane leave little room for an appreciable ``disk" dark matter (dDM) component. 

Measurements of the column densities leave more room to hide dDM, but are still consistent with no dDM at all.  Typical results for the total column density  include $\Sigma_{1.1\kpc} = 74 \pm 6~$M$_\odot/$pc${^2}$ \cite{holmberg:2004},   $\Sigma_{0.8\kpc} = 74^{+25}_{-12}~$M$_\odot/$pc${^2}$ \cite{siebert:2003} and $\Sigma_{1.1\kpc} = 71 \pm 6~$M$_\odot/$pc${^2}$ \cite{kuijken:1991}. On the other hand, different estimates for the total baryon column density range between 50 to 60~M$_\odot/$pc${^2}$ \cite{holmberg:2004,flynn:2006,bovy:2012,mckee:2015}.  Since $\rho_\hdm \sim 0.008~$M$_\odot/$pc${^3}$ corresponds to $ \Sigma_{\hdm,1.1\kpc} \sim 18 $~M$_\odot/$pc${^2}$, there is little room for additional dDM.

There are however two caveats. First, it was shown that kinematic determinations of the density at the MW plane suffer from systematic uncertainties due to the expected perturbation by spiral arm passages \cite{shaviv:2016}.  Because the density increase associated with the interstellar gas is abrupt, stars with a relatively small vertical oscillation ($\lesssim 100$~pc) cannot adjust ``adiabatically" to the changed potential such that the whole stellar distribution develops ``ringing" motion which can systematically distort the inferred mass density. The apparent contraction of the stars in the solar vicinity towards the plane is a signature of this effect \cite{shaviv:2016}. Without the constraint of ref.\ \cite{holmberg:2000}, a local disk of dark matter cannot be ruled. 

Moreover, estimates for the  total baryonic column density is obtained using a vertical potential which neglects the existence of excess dark matter in the disk. However, by introducing dark matter, the vertical potential is deeper such that the total baryonic column density inferred from observations and modeling is smaller, leaving more room for Dark Matter, as recently pointed out \cite{randall:2016}, and as borne also in the analysis below. 

With the above caveats considered, there is significant room for excess dark matter at the MW disk, with a column density of $\lesssim 20~$M$_\odot/$pc$^2$, as also pointed out by \cite{randall:2016}. It does not prove that a disk exists, since without reliable density measurement at the plane and sufficiently large uncertainties in the column densities, a no dDM solution is still possible. However, it becomes inconsistent once the paleoclimate data \cite{shaviv:2014} is considered. Below, we also show that other Massive Compact {\em Disk} Objects (``Macdos") are inconsistent implying that it cannot be an unseen baryonic component or gravitationally collapsed Dark Matter. 

 We begin in \S\ref{sec:model} with building a self-consistent model of the vertical structure of the MW. We continue in \S\ref{sec:stability} with a discussion of the disk stability to self gravity and in \S\ref{sec:implications} with the implications. In \S\ref{sec:alternative} we show that alternative explanations to the paleoclimatic data and the dDM are not plausible, and  then end with a discussion on the implications to dark matter  and a summary in  \S\ref{sec:discussion}. 
 
\section{Model for vertical structure}
\label{sec:model}
\secskip

We follow the standard methodology and approximation to solve for the vertical dependence of the various mass components and the gravitational potential, e.g., refs.\ \cite{holmberg:2000,mckee:2015}. We assume that each component follows a thermal equilibrium distribution of the form
$
\rho_i = \rho_{0,i} \exp \left( - {\Phi(z) / \sigma_{z,i}^2} \right).
$
We note however, like ref.\ \cite{randall:2016}, that some of the components have their local density determined from observations, while others, in particular the interstellar gas, have their column density determined. The values themselves are taken from ref.\ \cite{mckee:2015}. We also add to the model a standard hDM component with a constant background density of $0.008 \pm 0.005~$M$_\odot/$pc$^2$ \cite{bovy:2012} (i.e., we solve for 0.005, 0.008 and  $0.011~$M$_\odot/$pc$^2$). We also add a dDM component with a given $\rho_{0,\ddm}$ at the plane and vertical dispersion $\sigma_{z,\ddm}$.

For the gravitational potential we neglect the rotation curve term \cite{bovy:2012}, in which case we have 
\begin{equation}
	K_z = - {\partial \Phi \over \partial z},~~\mathrm{and}~~\Sigma(z) = \int_{-z}^{z} \rho(z') dz' = {\left| K_z \right| \over 2 \pi G}.
\end{equation}

To solve a model, the density at the MW plane is first guessed for the components with observational constraints on the column densities. The vertical galactic potential can then be integrated, giving the total column densities of all the different components, including those with a fixed column density. The  densities of the latter can then be iterated for until their integrated column density agrees with the observational constraints.

With a given model, we can plot the total baryonic and gravitating column densities (e.g., up to 1.1 kpc), the column density and dDM density at the Galactic plane, as depicted in fig.\,\ref{fig:05}.

\section{Disk Stability}
\label{sec:stability}
\secskip

In addition to the above observational considerations on the densities and column densities, it is interesting to analyze the stability of the dark matter disk to the effects of self gravity. Toomre  found a criterion to the instability of local axisymmetric disturbances of a thin disk of collisionless particles \cite{toomre:1964}, such as stars, which should pertain to dark matter particles as well. If we define $Q \equiv \sigma_R \kappa / 3.36 G \Sigma$, with $\sigma_U$ being the radial velocity dispersion and $\kappa$ the radial epicyclic frequency, then $Q<Q_{crit} = 1$ is a necessary condition for instability. The disk can be unstable to non-axisymmetric perturbations (i.e., to bars and spirals) for somewhat smaller densities, i.e., to $Q_{crit} \sim 1.2-1.5$. Disks which are unstable, with $Q<1$, will generate sufficient waves to kinematically heat up the disk, thus increasing $Q$ to ``stable" values. We therefore expect $Q \gtrsim 1$. 

Two primary complications arise when calculating $Q$ for the solar neighborhood. First, the local mass is heterogeneous. Not only is there an important contribution from gas, the stellar component can be described as a combination of different populations with different kinematic characteristics. The stability criterion can then be written as \cite{rafikov:2001} 
\begin{equation}
	2 \pi G k {\Sigma_g \over \kappa^2 + k^2 c_g^2}
	+ {2 \pi G k \over \kappa^2} \sum_{j=1}^{n} \Sigma_j \Psi_j > 1,
\end{equation}
with $c_g$ being the sound speed of the gas. Also,  
\begin{equation}
\Psi_j = {1 - \exp(-k^2 \sigma_j^2 / \kappa^2) I_0 (k^2 \sigma_j^2 / \kappa^2) \over k^2 \sigma_j^2 / \kappa^2},	
\end{equation}
and $I_0$ is the zeroth Bessel function.

The second modification is the effects of thick disks. This is important because one components' most unstable wavelength could be small compared with the scale height of another component. Although there is no exact solution to this problem, we can apply the useful reduction factor ansatz of ref.\ \cite{romeo:1992}, giving the criterion:
\begin{equation}
	2 \pi G k {\Sigma_g \over \kappa^2 + k^2 c_g^2}
	+ {2 \pi G k \over \kappa^2} \sum_{j=1}^{n} {\Sigma_j \Psi_j  \over 1+ k h_j} > 1,
\end{equation}
with $h_j = \Sigma/(2 \rho_{0,j})$ being the effective scale height of component $j$. Thus, given a model solution, we can calculate the value of $Q$. This is plotted in red in fig.~\ref{fig:05}. Evidently, dDM disks which are denser than about $0.1~$M$_\odot/$pc$^3$ at the plane are unstable. Note that we assume the radial and vertical dispersion of the dDM are the same. For stars, $\sigma_W \sim \sigma_U/2$ because dissipation heats the radial direction, which then leaks vertically. Although the ratio for a cooling disk is unclear, the dispersions should be comparable.
 
\section{Results and implications to dDM}
\label{sec:implications}
\secskip

\begin{figure*}[t]
\begin{minipage}[c]{0.67\textwidth}
\includegraphics[width=\textwidth]{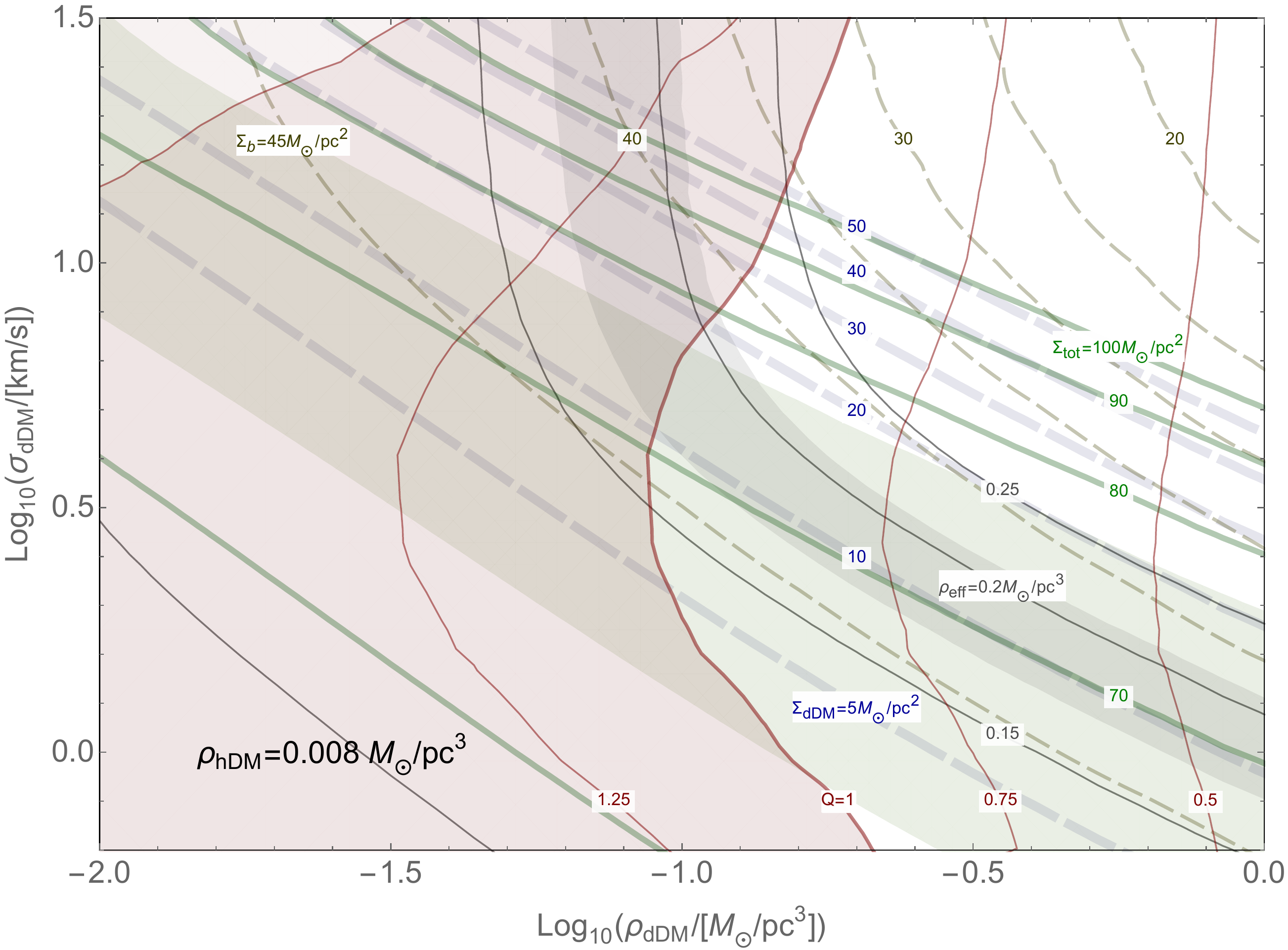}	
\end{minipage}
\begin{minipage}[c]{0.322\textwidth}
\includegraphics[width=\textwidth]{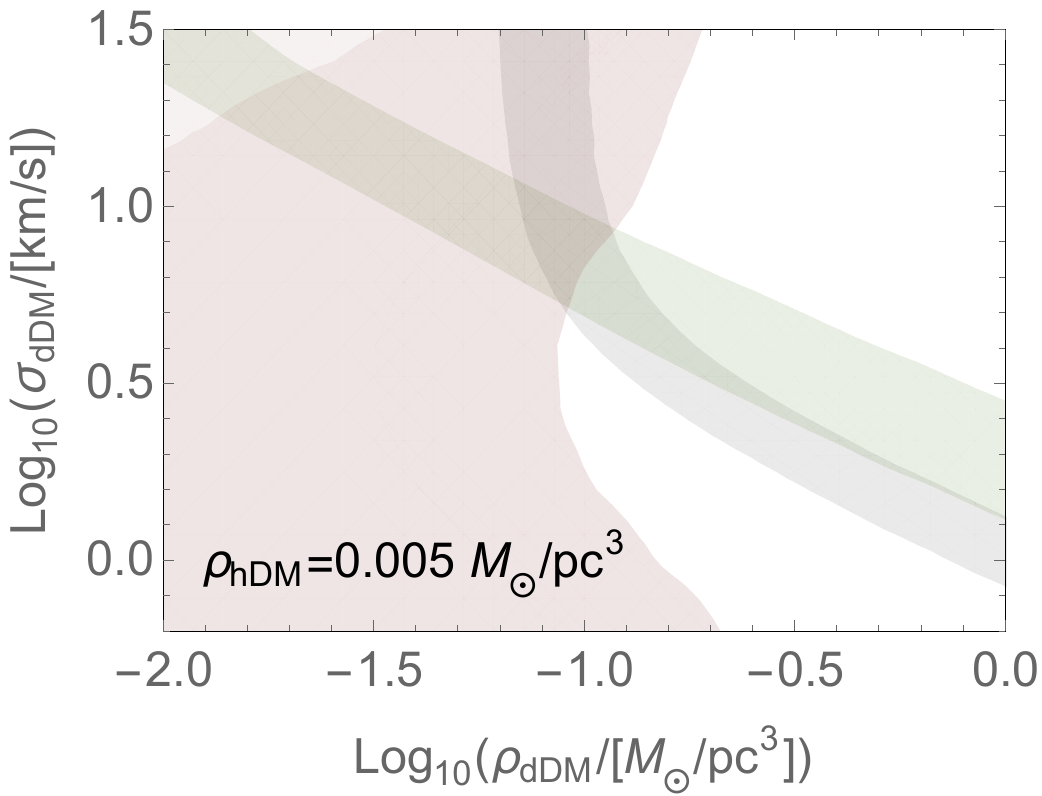}	
\includegraphics[width=\textwidth]{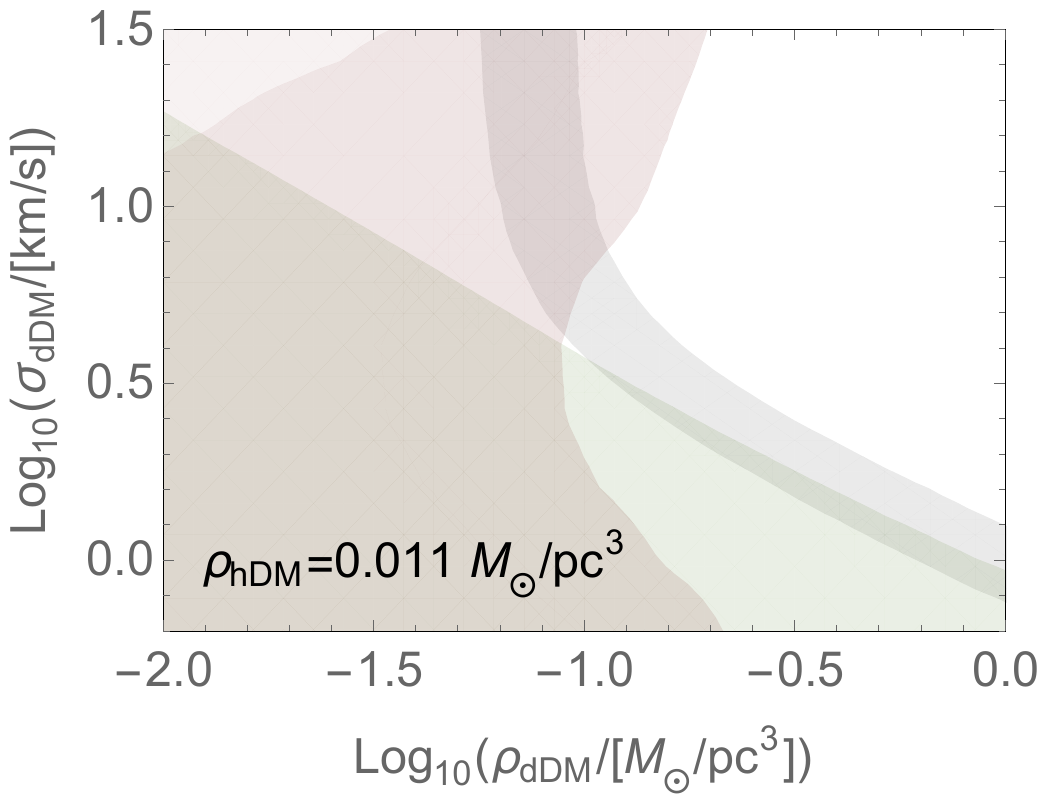}	
\end{minipage}
\vskip -6pt
\caption{\label{fig:05} Model solutions assuming $\rho_{\hdm} = 0.008~$M$_\odot/$pc$^3$ (large panel), $\rho_{\hdm} = 0.005~$M$_\odot/$pc$^3$ (top right) and $\rho_{\hdm} = 0.011~$M$_\odot/$pc$^3$ (bottom right). Large panel includes contour levels of $\Sigma_b$ (dashed gray), $\Sigma_{tot}$ (solid green, with the green shaded region denoting observational constraints),  $\Sigma_{\ddm}$ (dashed blue) and the equivalent constant density $\rho_\mathrm{eff}$ needed to give the observed 32 Mr oscillation seen in the geological data (solid gray, and shaded region denoting the paleoclimatic constraint) are given as a function of the $\rho_{0,\ddm}$ and $\sigma_{\ddm}$. The red contours denote the Toomre $Q$ value. A region near $\rho_{0,\ddm} \sim 0.1 ~$M$_\odot/$pc$^3$  and $\sigma_{\ddm} \sim 6 $\,km$/$s satisfies all constraints. The small plots are abridged and only include observationally constrained regions. }
\end{figure*}

The model results are plotted in fig.\ \ref{fig:05}. The different constraints are denoted by the shaded regions. For the total mass we take a range of $71\pm 6$~M$_\odot/$pc$^2$ as determined by ref.\ \cite{holmberg:2004}. For the effective density we take the paleoclimate determinations \cite{shaviv:2014}. Several consequences can be reached. First, a no dDM solution is permissible only if the background density of hDM is on the large side and the paleoclimate data is discarded. If it is accepted, however, then only hDM densities of $\lesssim 0.01$M$_\odot/$pc$^3$ can provide solutions satisfying all the constraints, which span $\rho_\mathrm{\ddm} = 0.11 \pm 0.03$~M$_{\odot}/$pc$^3$.  It also has a surface density $\Sigma_{\ddm} = 15 \pm 5$~M$_{\odot}/$pc$^2$ and a vertical velocity dispersion of $\sigma_{W,\ddm} = 8.0 \pm 4.5$~km/s. Namely, if the oscillation in the paleoclimate data is a signature of the  vertical motion of the solar system, then an additional kinematically cold component in necessarily present in the disk. 

\section{Alternative explanations?}
\label{sec:alternative}
\secskip

The conclusion that dDM exists rests on several assumptions. As mentioned above, the first is that the paleoclimate data is due to the vertical motion of the solar system. It requires that the discrepancy between the locally measured baryon density and the effective total density measured over 550 Ma is not due to large density variations. Last, it assumes that the unexplained component is non-baryonic and not, for example, massive compact disk objects of baryonic origin. 

The statistical significance of the paleoclimate signal was discussed in ref.\ \cite{shaviv:2014}. It is clear beyond any doubt that the periodic signal exists (at 17$\sigma$). If it is not due to the vertical motion it must be some other regular signal (e.g., due to some unknown very long interaction in the planetary interactions), which coincidentally has the correct phase to be the vertical motion (1 in 6 probability) and a secondary frequency modulation with a correct phase and period to mimic the radial epicyclic motion of the solar system (1 in 60 probability). 

\begin{figure}
\includegraphics[width=3.3in]{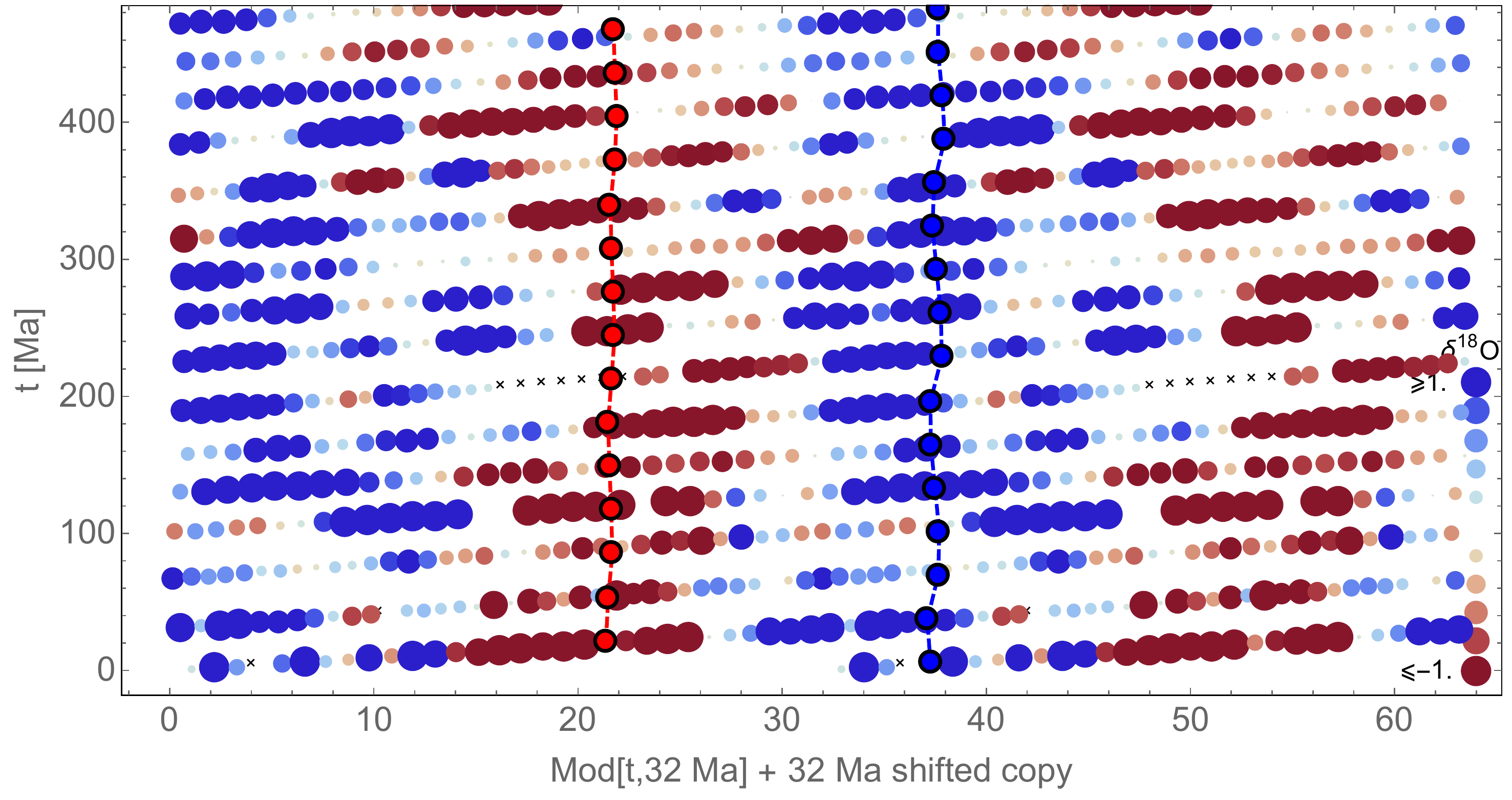}
\vskip -0pt
\caption{\label{fig:raster} The same as fig.\ 4 of ref.\ \cite{shaviv:2014}, except that time is distorted to remove the radial epicyclic motion described in the text. The vertical axis spans the Phanerozoic. The horizontal axis is the modified time folded over a 32 Ma period. For convenience, the horizontal axis shows two 32 Ma periods. The blue and red circles (connected by dashed lines) are the modeled  plane crossings (blue) and the maximal excursions from the plane (red), respectively. The disk radii and color correspond to the detrended and high pass filtered $\delta^{18}$O signal, as given by the scale on the right (in $\permil$).
 }\vskip -12pt
\end{figure}

If the baryon / effective density inconsistency is due to density variations, then the average ISM gas density has to be large by a factor of a few more than the local density. However, such large variations in the density (of order a factor of 2) should leave a fingerprint in the paleoclimate data in the form of cycle to cycle jumps that are of order $1/\sqrt{2}$, or about 20 Ma. Fig.\ \ref{fig:raster} replots  fig.\ 4 of ref.\ \cite{shaviv:2014}, which is the detrended paleotemperature  proxy data folded over the 32 Ma period. The difference is that time is here distorted to remove the radial epicyclic motion by defining a ``distorted" time $\tilde{t}$ as $d\tilde{t} = dt /\sqrt{ \rho(R(t)) / \rho_0 }$, with $\rho(R(t))$ being the modeled density in the plane at the modeled $R(t)$, due to the radial epicyclic motion. Since some of the apparent variations can be due to aging errors (whether measurement or additional climate variations), some of the typically $\lesssim 5$~Ma cycle to cycle jumps can be due changes in the density. Thus, the density variations are at most $\delta \rho / \rho \lesssim 0.3$, which cannot explain the large discrepancy between baryon and paleoclimate measurements. 

Although measurement of the missing gravitational component does not provide any direct indication to its nature, the fact that it can keep itself kinematically cold implies that it can self interact. This is apparent from the small vertical velocity dispersion, of $8.0 \pm 4.5$~km/s, compared with the typically $25$ to $30$~km/s dispersion of stars older than 10 Gyrs \cite{nordstrom:2008}. Thus, not only is it implausible for compact baryonic DM to explain the discrepancy, it would also imply that the dDM cannot consist of gravitationally collapsed DM objects, i.e., ``dark stars".

\section{Discussion and Summary}
\label{sec:discussion}
\secskip

The main argument against the existence of a dDM component is the small difference between the  measured baryonic density and kinematic determination of the local mass density. However, spiral arm passages distort the inferred mass density by ${\cal O}(1)$. On the other hand, paleoclimate data indicates that the local mass density is about twice larger than the local baryonic matter, implying that a cooling dDM component should be present, since it cannot be consistently explained otherwise.  The fact that the dDM cools down to form a disk is relevant for two major reasons.

First, if a cooling dDM component exists and can overcome the heating from viscous stirring, then the prediction should be that it cooled down to a velocity dispersion which is marginally stable, i.e., Toomre's $Q \gtrsim 1$. Below it, the disk would be unstable and heat itself up by exciting significant waves.  Since the $Q \sim 1$ line is roughly vertical in the dDM velocity dispersion density plane (see fig.\ \ref{fig:05}), any dDM component comparable to the baryonic one, if it exists and can cool, should have a density $\rho_{0} \sim 0.1$~M/pc$^3$. Namely, the paleoclimatic measurement recovers the theoretical prediction.

Allowing the dDM to cool requires that the cooling reaction rate is several time faster than the Hubble rate. For example, we can consider cooling through a reaction $2d \rightarrow 2d+\ell$, where $d$ is a dDM particle and $\ell$ is a light DM particle required to take the kinetic energy, then $	n \sigma_{dd\ell} v \gtrsim {\cal N} H $ where $n$ is the number density of dDM before the disk cools down (assuming it is formed ``puffed"), $v$ is the typical Keplerian velocity which characterizes the typical random component that a puffed up dDM halo would have. ${\cal N}$ is the typical number of interactions required for the cooling to take place. Taking $v \approx \sqrt{G M_\mathrm{MW\odot}/r_\odot}$ with $M_\mathrm{MW\odot}$ the amount of mass within our galactic radius $r_\odot$, then one finds that dDM at $r_\odot$ can cool if 
\begin{equation}
{	\sigma_{dd\ell} \over m_d} \gtrsim {4 \pi {\cal N} \over 3 \alpha}  {H r_\odot^{7/2} \over G^{1/2} M_{MW\odot}^{3/2}} \approx {{\cal N} \over \alpha} 0.03 {cm^2 \over gr}.
\end{equation}
Here we assumed that a fraction $\alpha$ of the DM mass is in the cooling component. For $\alpha \sim 0.1$ and ${\cal N} \sim 10$ we find that the cross section should satisfy ${\sigma_{dd / m_H}} \gtrsim 3~$cm$^2/$g. 

Another cooling reaction could be inverse-Compton like cooling, through $d+  \ell \rightarrow d+ \ell$. However it may require a relic $\ell$ background which on one hand has to be cold enough as to not leave a Baryon-like Acoustic Oscillation in the cosmic microwave background \cite{cyr:2014}, but not too cold to leave a negligible background on which the $dDM$ cannot cool. 

Last, we note that a kinematically cold and dense disk, which could periodically perturb the Oort cloud (and cause mass extinctions) is also ruled out as it would be kinematically unstable. It would develop horizontal perturbations which would quickly heat the disk. It cannot form collapsed objects (``dark stars"), as those will then have a Hubble time to heat to $\gtrsim 25$~km/s.

Many thanks Yoram Lithwick, Erik Kuflik, Yonit Hochberg, Kris Sigurdson, James Owen, and Scott Tremaine for fruitful discussions. This research project was supported by the
I-CORE Program of the Planning and Budgeting Committee and the Israel Science Foundation (center 1829/12) and by ISF grant no.\ 1423/15.

\def\apjl{ApJ}
\def\aap{Astron.\ Astrophys.}
\def\bain{Bull.\ Astron.\ Inst.\ Netherlands}
\def\mnras{MNRAS}
\def\MNRAS{MNRAS}

\bibliography{dDM.bib}

\end{document}